\documentclass{aa}

\usepackage[varg]{txfonts}
\usepackage[colorlinks=true, linkcolor=blue, citecolor=violet, urlcolor=gray]{hyperref}
\usepackage{xcolor}
\usepackage{listings}

\bibpunct{(}{)}{;}{a}{}{,}

\defcitealias{2012MNRAS.421.1007K}{K12}
\defcitealias{2014MNRAS.441.1340V}{V14}

\begin{document}

\title{The TNG50-SKIRT Atlas: \\
wavelength dependence of the effective radius}

\titlerunning{TNG50-SKIRT Atlas}

\author{%
Maarten Baes\inst{\ref{UGent}}
\and
Aleksandr Mosenkov\inst{\ref{BYU}}
\and
Raymond Kelly\inst{\ref{BYU}}
\and
Abdurro'uf\inst{\ref{JHU},\ref{STScI}}
\and
Nick Andreadis\inst{\ref{UGent}}
\and
Sena Bokona Tulu\inst{\ref{UGent},\ref{Jimma}}
\and
\\
Peter Camps\inst{\ref{UGent}}
\and 
Abdissa Tassama Emana\inst{\ref{UGent},\ref{Jimma}}
\and
Jacopo Fritz\inst{\ref{UNAM}}
\and
Andrea Gebek\inst{\ref{UGent}}
\and
Inja Kova{\v{c}}i{\'{c}}\inst{\ref{UGent}}
\and
Antonio La Marca\inst{\ref{SRON},\ref{Kapteyn}}
\and
\\
Marco Martorano\inst{\ref{UGent}}
\and
Angelos Nersesian\inst{\ref{UGent}}
\and
Vicente Rodriguez-Gomez\inst{\ref{UNAM}}
\and
Crescenzo Tortora\inst{\ref{Napoli}}
\and
\\
Ana Tr{\v{c}}ka\inst{\ref{UGent}}
\and
Bert Vander Meulen\inst{\ref{UGent}}
\and
Arjen van der Wel\inst{\ref{UGent}}
\and
Lingyu Wang\inst{\ref{SRON},\ref{Kapteyn}}
}

\institute{%
Sterrenkundig Observatorium, Universiteit Gent, Krijgslaan 281 S9, B-9000 Gent, Belgium\\ \email{maarten.baes@ugent.be}
\label{UGent}
\and
Department of Physics and Astronomy, N283 ESC, Brigham Young University, Provo, UT 84602, USA
\label{BYU}
\and
Center for Astrophysical Sciences, Department of Physics and Astronomy, The Johns Hopkins University, 3400 N Charles St., Baltimore, MD 21218, USA
\label{JHU}
\and
Space Telescope Science Institute, 3700 San Martin Drive, Baltimore, MD 21218, USA
\label{STScI}
\and
Physics Department, College of Natural Sciences, Jimma University, PO Box 378, Jimma, Ethiopia
\label{Jimma}
\and
Instituto de Radioastronom{\'{\i}}a y Astrof{\'{\i}}sica, Universidad Nacional Aut{\'{o}}noma de M{\'{e}}xico, Morelia, Michoac{\'{a}}n 58089, Mexico
\label{UNAM}
\and
SRON Netherlands Institute for Space Research, Landleven 12, 9747 AD, Groningen, The Netherlands
\label{SRON}
\and
Kapteyn Astronomical Institute, University of Groningen, Postbus 800, 9700 AV, Groningen, The Netherlands
\label{Kapteyn}
\and
INAF -- Osservatorio Astronomico di Capodimonte, Salita Moiariello 16, I-80131 Napoli, Italy
\label{Napoli}
}

\authorrunning{M. Baes et al.}

\date{\today}

\abstract{%
Galaxy sizes correlate with many other important properties of galaxies, and the cosmic evolution of galaxy sizes is an important observational diagnostic for constraining galaxy evolution models. The effective radius is probably the most widely used indicator of galaxy size. We used the TNG50-SKIRT Atlas to investigate the wavelength dependence of the effective radius of galaxies at optical and near-infrared (NIR) wavelengths. We find that, on average, the effective radius in every band exceeds the stellar mass effective radius, and that this excess systematically decreases with increasing wavelength.  The optical {\textit{g}}-band (NIR ${\textit{K}}_{\text{s}}$-band) effective radius is on average 58\% (13\%) larger than the stellar mass effective radius. Effective radii measured from dust-obscured images are systematically larger than those measured from dust-free images, although the effect is limited (8.7\% in the {\textit{g}}-band, 2.1\% in the ${\textit{K}}_{\text{s}}$-band). We find that stellar population gradients are the dominant factor (about 80\%) in driving the wavelength dependence of the effective radius, and that differential dust attenuation is a secondary factor (20\%). Comparing our results to recent observational data, we find offsets in the absolute values of the median effective radii, up to 50\% for the population of blue galaxies. We find better agreement in the slope of the wavelength dependence of the effective radius, with red galaxies having a slightly steeper slope than green--blue galaxies. Comparing our effective radii with those of galaxies from the Siena Galaxy Atlas in separate bins in ${\textit{z}}$-band absolute magnitude and ${\textit{g}}-{\textit{z}}$ colour, we find excellent agreement for the reddest galaxies, but again significant offsets for the blue populations: up to 70\% for galaxies around $M_z=-21.5$. This difference in median effective radius for the bluer galaxies is most probably due to intrinsic differences in the morphological structure of observed and TNG50 simulated galaxies. Finally, we find that the median effective radius in any broadband filter increases systematically with decreasing ${\textit{u}} - {\textit{r}}$ colour and with increasing galaxy stellar mass, total SFR, sSFR, and dust-to-stellar-mass ratio. For the slope of the wavelength dependence of $R_{\text{e}}$, however, there does not seem to be a systematic, monotonic correlation with any of these global properties.
}


\maketitle


\section{Introduction}

Of all the morphological characteristics of a galaxy, size is arguably the most important one. Galaxy sizes correlate with many other important properties of galaxies, such as stellar mass, surface brightness, and stellar density \citep{1977ApJ...217..406K, 1994ARA&A..32..115R, 2003MNRAS.343..978S, 2014ApJ...788...28V}. The cosmic evolution of galaxy sizes is an important observational diagnostic for constraining galaxy evolution models \citep{2006ApJ...650...18T, 2007MNRAS.382..109T, 2008ApJ...688...48V, 2014ApJ...788...28V, 2018MNRAS.480.1057R, 2023ApJ...946...71C}. Galaxy sizes are also important to define central dark matter fractions, and  can be used together with these dark matter fractions to test galaxy evolution models \citep[e.g.][]{2018MNRAS.473..969T}.

There are different ways to characterise the size of a galaxy \citep{1976ApJ...209L...1P, 1980ApJS...43..305K, 2019PASA...36...35G, 2020MNRAS.493...87T}. The effective or half-light radius is probably the most widely used one. One of the main advantages of the effective radius as an indicator of galaxy size is its robustness against many observational issues, such as the depth of images \citep[e.g.][]{2001MNRAS.326..869T}. On the other hand, the effective radius is an on-sky property and not an intrinsic 3D property of a galaxy, and its value is expected to depend on the orientation and the band in which we observe the galaxy \citep{2006A&A...456..941M, 2008MNRAS.388.1708G, 2013A&A...557A.137P, 2017MNRAS.468L..31D}. 

Several teams demonstrated observationally that the effective radius is indeed wavelength-dependent. \citet{2005JKAS...38..149K} and \citet{2010MNRAS.408.1313L} found a 30\% decrease in the effective radius of early-type galaxies when moving from the optical {\it{g}}- or {\it{V}}-bands to the near-infrared (NIR) {\it{K}}-band. \citet{2012MNRAS.421.1007K} and \citet{2014MNRAS.441.1340V} extended these studies by including late-type galaxies, and found an overall $\sim$30\% decrease in effective radius when moving from {\it{g}}- to {\it{K}}-band values, with the decrease slightly depending on galaxy colour or S\'ersic index. Other studies indicating a similar dependence of effective radius on wavelength include \citet{2015MNRAS.447.2603L}, \citet{2015MNRAS.454..806K, 2016MNRAS.460.3458K}, \citet{2017A&A...605A..18C}, \citet{2018MNRAS.478.5410D}, \citet{2020A&A...633A.104P}, \citet{2020A&A...641A.119B}, and \citet{2023A&A...673A..63N}. Advanced studies considering  multi-wavelength bulge-disc composition of large galaxy samples include \citet{2022A&A...664A..92H} and \citet{2022MNRAS.516..942C}.

The systematic wavelength dependence is generally understood to be a combination of two distinct effects: stellar population gradients and differential dust attenuation. Both early-type and late-type galaxies generally show negative metallicity and age gradients \citep{2010MNRAS.407..144T, 2014A&A...570A...6S, 2016MNRAS.463.2799I, 2017A&A...608A..27G, 2017MNRAS.465.4572Z, 2018A&A...615A..27L, 2019MNRAS.483.1862Z, 2020AJ....159..195D, 2021MNRAS.502.5508P, 2023A&A...673A.147P}. These stellar population gradients naturally lead to colour gradients, and thus a systematic decrease in the half-light radius for increasing wavelength. On the other hand, dust attenuation primarily obscures the central regions of galaxies, as evidenced from spatially resolved imaging of galaxies at far-infrared wavelengths \citep{2009ApJ...701.1965M, 2010A&A...518L..72P, 2015A&A...581A.103G, 2017A&A...605A..18C, 2020MNRAS.495.2305G} and studies of overlapping galaxies \citep{2001AJ....121.1442K, 2001AJ....122.1369K, 2009AJ....137.3000H}. This differential dust attenuation leads to a flattening of the surface brightness profile, and hence to an increase in the half-light radius. Dust attenuation, being strongly wavelength-dependent \citep{1990ARA&A..28...37M, 2003ARA&A..41..241D, 2005MNRAS.360.1413B}, automatically generates a decrease in half-light radius with increasing wavelength. 

These two effects, stellar population gradients and differential dust attenuation, are hard to disentangle for observed galaxies. They can be disentangled, however, for simulated galaxies. In this work we use the TNG50-SKIRT Atlas \citep[TSA,][]{Baes2024a}, a large multi-wavelength synthetic image atlas based on simulated galaxies extracted from the TNG50 cosmological hydrodynamics simulation \citep{2019MNRAS.490.3196P, 2019MNRAS.490.3234N}, to investigate the origin of the wavelength dependence of the effective radius in the optical--NIR wavelength domain. Since dust attenuation can artificially be turned on and off, the relative contribution of the two contributors can be determined. Disentangling these two effects is the main goal of this work.

This paper is organised as follows. In Sect.~{\ref{Methods.sec}} we discuss the data we use and the methods we adopt to determine the effective radius. In Sect.~{\ref{Results.sec}} we present the results of our analysis. More specifically we show how the effective radius of the simulated galaxies in our sample depends on wavelength and we discuss the origin of this wavelength dependence. In Sect.~{\ref{Discussion.sec}} we discuss our findings: we compare our results to observations and we investigate how the effective radius and its wavelength dependence depend on global galaxy parameters. We present a summary and our conclusions in Sect.~{\ref{Conclusions.sec}}.


\section{Sample and methods}
\label{Methods.sec}

In this section we describe the sample of simulated galaxies we use for our analysis (Sect.~{\ref{TSA.sec}}) and the methods we adopt to determine effective radii (Sect.~{\ref{measurement.sec}}) and half-mass radii (Sect.~{\ref{rhalf.sec}}) for each simulated galaxy.

\begin{figure*}
\centering
\includegraphics[width=\textwidth]{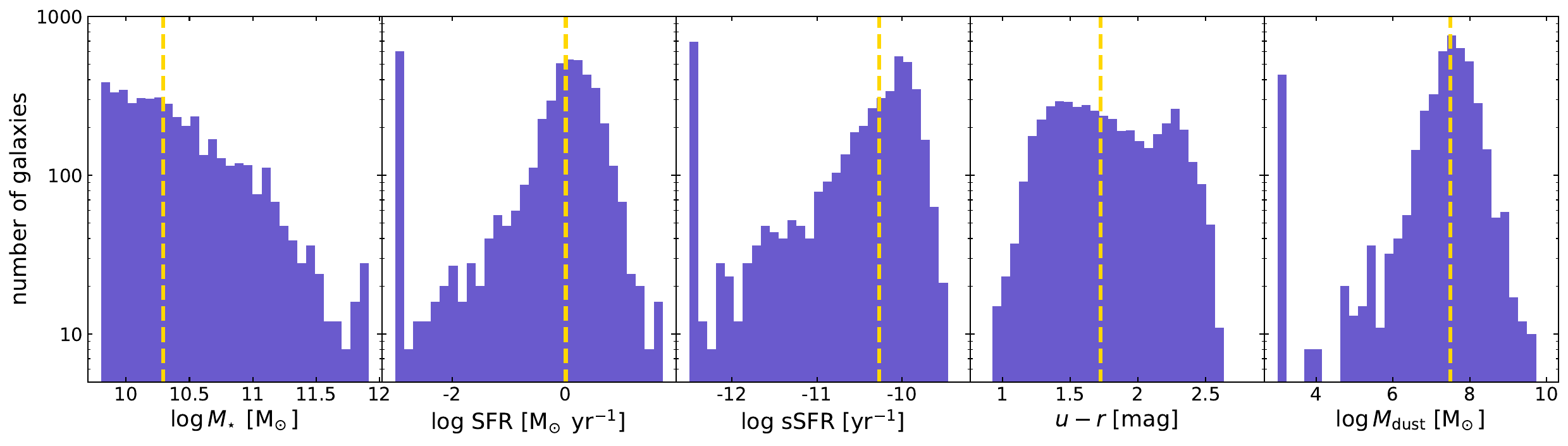}
\caption{Histograms of a number of global properties of the galaxies in our sample: stellar mass, star formation rate, specific star formation rate, ${\textit{u}}-{\textit{r}}$ colour, and dust mass. The dashed lines indicate the median values of the sample. The peaks at low values in the SFR and sSFR histograms correspond to galaxies without ongoing star formation. Similarly, the peak at low dust mass corresponds to galaxies without interstellar dust content.}
\label{histograms_sample.fig}
\end{figure*}

\begin{figure*}
\centering
\includegraphics[width=0.8\textwidth]{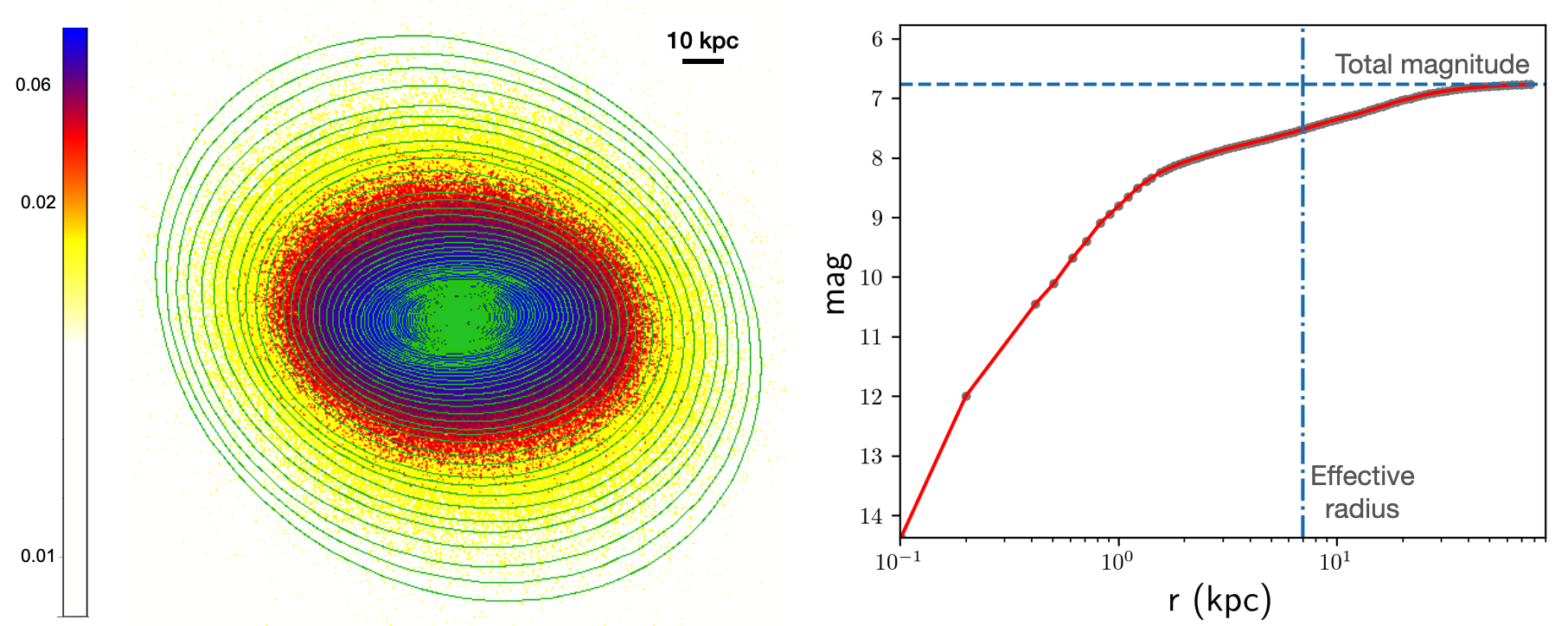}
\caption{Illustration of the curve-of-growth analysis described in Sect.~{\ref{measurement.sec}}, here applied to the 2MASS {\it{H}}-band image of simulated galaxy TNG\,000001 in orientation O1. The 2D surface brightness distribution is fitted with a series of concentric ellipses, and the total integrated flux within each ellipse is determined. The effective radius is the semi-major axis of the ellipse that contains half of the total flux.}
\label{growth_curve_ellipses.fig}
\end{figure*}

\subsection{The TNG50-SKIRT Atlas}
\label{TSA.sec}

The TSA \citep{Baes2024a} is a synthetic image atlas for a complete stellar-mass selected sample of 1154 galaxies extracted from the TNG50 simulation \citep{2019MNRAS.490.3196P, 2019MNRAS.490.3234N}. TNG50 is the highest-resolution version of the IllustrisTNG suite of cosmological hydrodynamical simulations \citep{2018MNRAS.480.5113M, 2018MNRAS.477.1206N, 2018MNRAS.475..624N, 2018MNRAS.475..648P, 2018MNRAS.475..676S}. The simulation has a baryonic mass resolution of $8.5\times10^4~{\text{M}}_\odot$ and a spatial resolution of 70--140 pc. 

We selected all 1154 galaxies with total stellar mass between $10^{9.8}$ and $10^{12}~{\text{M}}_\odot$ from the $z=0$ snapshot of the TNG50 simulation. For each of these simulated galaxies, the TSA contains high-resolution images in 18 broadband filters covering the UV to NIR wavelength range\footnote{More specifically, we generated images in the GALEX FUV- and NUV-bands, the Johnson {\textit{UBVRI}}-bands, the LSST {\textit{ugrizy}}-bands, the 2MASS {\textit{JHK}}$_{\text{s}}$-bands, and the WISE W1- and W2-bands.}. All images are generated with the SKIRT radiative transfer code \citep{2015A&C.....9...20C, 2020A&C....3100381C} and account for different stellar populations and absorption and scattering by interstellar dust in a realistic 3D setting. The SKIRT post-processing simulations were based on the prescriptions by \citet{2016MNRAS.462.1057C, 2018ApJS..234...20C, 2022MNRAS.512.2728C}, \citet{2021MNRAS.506.5703K}, and \citet{2022MNRAS.516.3728T}. For each galaxy, the image atlas contains images for five different random observer positions (with observing positions O4 and O5 antipodal). In addition to the dust-obscured images, the atlas contains, for each observer position, synthetic dust-free images in all 18 bands, and stellar mass surface density, mean stellar age, mean stellar metallicity, and dust mass surface density maps. All images and parameter maps contain 1600~$\times$~1600 pixels with a pixel scale of 100 pc, corresponding to a field of view of 160~kpc. 

The TSA sample contains 869 star-forming and 285 quiescent galaxies when the value ${\text{sSFR}} = 10^{-11}~{\text{yr}}^{-1}$ is used as the dividing line between the two populations. Figure~{\ref{histograms_sample.fig}} shows histograms of a number of physical characteristics of the sample. The stellar masses, star formation rates (SFR), and specific star formation rates (sSFR) are directly taken from the TNG50 public database \citep{2019ComAC...6....2N}. The ${\textit{u}}-{\textit{r}}$ colours are calculated from the SKIRT-generated maps, and take into account the effects of dust attenuation. The dust masses are calculated by integrating the dust mass surface density maps. We note that the TNG50 hydrodynamical simulations do not explicitly contain a prescription for interstellar dust, so dust was added to the galaxies in post-processing using the assumption that the dust density is proportional to the gas metallicity in cold, dense gas \citep[for details, see][]{2022MNRAS.516.3728T, Baes2024a}.

The distribution of stellar masses follows the typical Schechter distribution with a strong decline for increasing stellar mass. While our sample consists of all galaxies with stellar masses between $10^{9.8}$ and $10^{12}~{\text{M}}_\odot$, the median stellar mass is only $10^{10.3}~{\text{M}}_\odot$. The sample contains galaxies with a variety in SFR, ranging from totally quiescent galaxies to actively star-forming galaxies with ${\text{SFR}} > 20~{\text{M}}_\odot~{\text{yr}}^{-1}$ and ${\text{sSFR}} > 10^{-9.5}~{\text{yr}}^{-1}$.  The median values for the sample are $\langle {\text{SFR}} \rangle = 1~{\text{M}}_\odot~{\text{yr}}^{-1}$ and $\langle {\text{sSFR}} \rangle = 10^{-10.3}~{\text{yr}}^{-1}$, indicating that most of the galaxies in our sample are star-forming. The distribution of ${\textit{u}} - {\textit{r}}$ colour covers the range between 0.9 and 2.7 and is bimodal, as also noted in observed samples of galaxies in the local Universe \citep{2001AJ....122.1861S, 2004ApJ...600..681B, 2004ApJ...615L.101B}. The median value is 1.72. Finally, the distribution of dust mass is clearly peaked with a long tail towards low dust masses and with a median dust mass of $10^{7.5}~{\text{M}}_\odot$. Such a distribution is consistent with the typical dust masses found in nearby galaxies \citep{2013MNRAS.428.1880A, 2014A&A...565A.128C, 2019A&A...624A..80N, 2020ApJ...889..150A, 2021A&A...649A..18G}.

\subsection{Determination of the effective radii}
\label{measurement.sec}

\begin{figure*}
\centering
\includegraphics[width=0.85\textwidth]{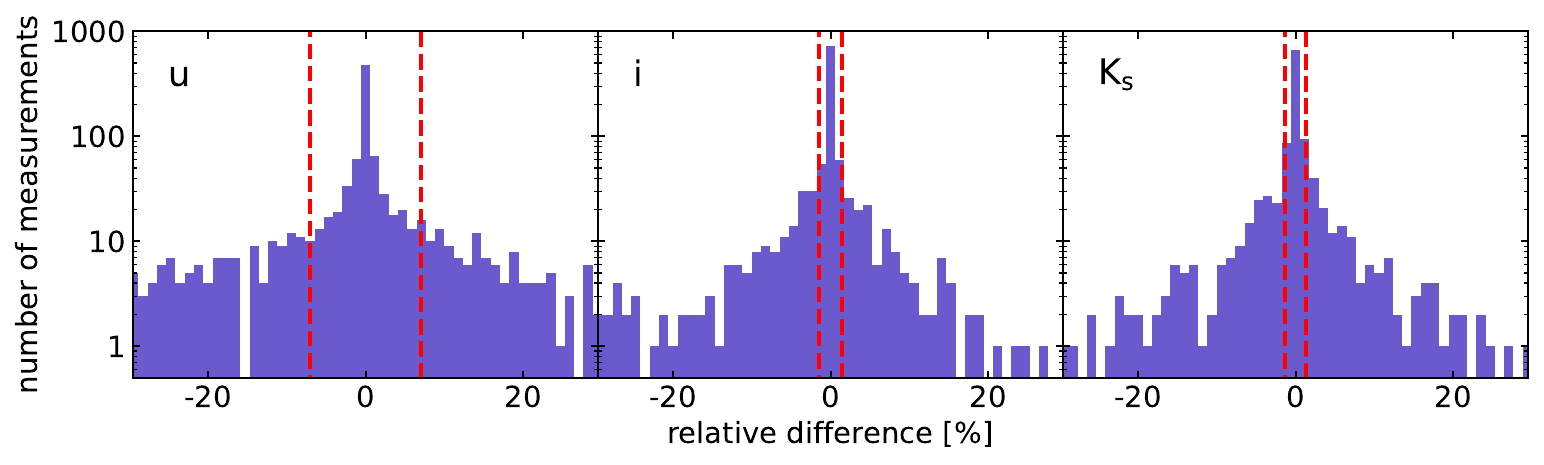}
\caption{Histograms of the relative difference between the effective radius measured from the antipodal O4 and O5 observer positions in the LSST {\textit{u}}- and {\textit{i}}-bands and the 2MASS ${\textit{K}}_{\text{s}}$-band. The dashed red lines represent the 16\% and 86\% quantiles of the distribution. Half of the width of this confidence interval can be used as an upper limit for the relative uncertainty on our $R_{\text{e}}$ measurements.}
\label{relerr_Re.fig}
\end{figure*}

We measured the effective radius in all LSST and 2MASS wavebands using a curve-of-growth analysis, similar to the approach applied to the galaxies from the Siena Galaxy Atlas by \citet{2023ApJS..269....3M}. Following most observational studies of the wavelength dependence of the effective radius \citep[e.g.][]{2005JKAS...38..149K, 2010MNRAS.408.1313L, 2012MNRAS.421.1007K, 2014MNRAS.441.1340V, 2015MNRAS.447.2603L}, we focus on the optical and NIR regime, from the ${\textit{u}}$- to the ${\textit{K}}_{\text{s}}$-band. 

As a first step we first convolve each SKIRT post-processed galaxy image with a Gaussian kernel with FWHM of 5 pixels. This step was taken to smooth out fine details and Monte Carlo noise in our galaxy images. We experimented with different values of the smoothing lengths and settled on 5~{\text{pix}} as a trade-off between a significant degradation of the resolution and a sufficient noise reduction. The second step consisted of fitting elliptical isophotes to the individual images by means of the standard IRAF STSDAS/ELLIPSE routine \citep{1987MNRAS.226..747J}, with the galaxy centre fixed. We determined the total integrated flux within each elliptical aperture, and plotting them as a function of increasing semi-major axis yielded the curve-of-growth. This curve was approximated with a cubic spline fit from which the half-light radius was measured. This procedure was independently executed for every single image in our atlas. Figure~{\ref{growth_curve_ellipses.fig}} illustrates our approach.

For a limited number of images this method did not converge, either because the galaxies were extremely asymmetrical, or because there were technical issues (for example, for galaxies located very close to the border of the simulation volume). We discarded all measurements for a given galaxy and observer position if the ellipse fitting routine failed for at least one band. In total, we had to discard 129 sets of images out of a total of $5\times1154 = 5770$ sets, corresponding to a 2.2\% rejection rate.

To characterise the accuracy of our effective radius measurements, we compared the values obtained independently for the antipodal O4 and O5 observer positions (the images corresponding to observer position O5 were only used as a check for the accuracy of our analysis but were not used in the statistical analysis in the remainder of this work). Figure~{\ref{relerr_Re.fig}} shows the histograms of the relative error between the two measurements for the {\textit{u}}-, {\textit{i}}-, and ${\textit{K}}_{\text{s}}$-bands. The relative difference between the O4 and O5 measurements decreases with increasing wavelength, which can be attributed to the diminishing effect of dust attenuation and the smoother appearance of galaxies at longer wavelengths. For each band, we calculated the 16\% and 84\% quantiles of the distribution of the relative difference and considered half of the width of this confidence interval as an upper limit for the relative uncertainty on our $R_{\text{e}}$ measurements, as the O4 and O5 images are not perfectly mirrored versions due to different levels of dust attenuation. This value decreases from 7.0\% in the {\it{u}}-band to 1.3\% in the ${\textit{K}}_{\text{s}}$-band. 

\subsection{Determination of half-mass radii}
\label{rhalf.sec}

\begin{figure}
\centering
\includegraphics[width=0.9\columnwidth]{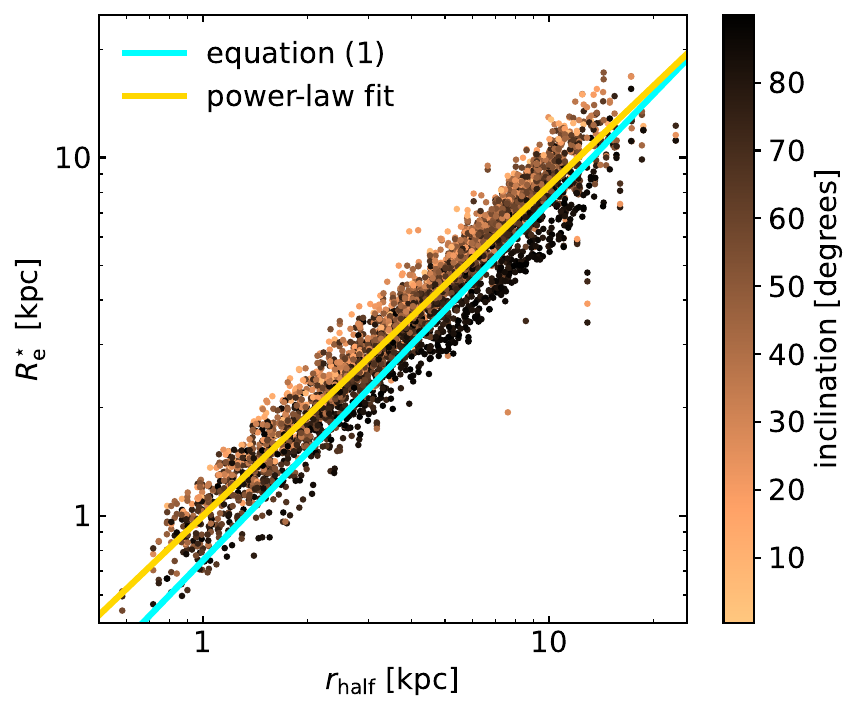}
\caption{Correlation between the 3D particle half-mass radius $r_{\text{half}}$ and the half-mass radius $R_{\text{e}}^\star$ determined as the semi-major axis of the elliptical iso-density contour that contains half of the total projected stellar mass. Each galaxy is represented by four dots, one for each of the four independent observer's positions. The solid yellow line is the best-fitting power-law relation.}
\label{halfmassradius.fig}
\end{figure}

We determined the effective radius in the stellar mass surface density maps in the same way as for the synthetic images. The corresponding effective radius, denoted as $R_{\text{e}}^\star$, is the semi-major axis of the elliptical iso-density contour that contains half of the total projected stellar mass. In Fig.~{\ref{halfmassradius.fig}} we compare $R_{\text{e}}^\star$ with the half-mass radius $r_{\text{half}}$ obtained directly from the TNG50 stellar particle data as the radius of the sphere that contains half of the total stellar mass. For completely spherical models with realistic surface density profiles, we expect a relation
\begin{equation}
R_{\text{e}}^\star \approx \frac34\,r_{\text{half}},
\label{Rerhalf}
\end{equation}
as demonstrated by \citet{1991A&A...249...99C} and \citet{2010MNRAS.406.1220W}. 

Fig.~{\ref{halfmassradius.fig}} shows the correlation between $r_{\text{half}}$ and $R_{\text{e}}^\star$ for all galaxies in our image atlas. Every galaxy has, by definition, a single value for $r_{\text{half}}$ and four different values for $R_{\text{e}}^\star$, corresponding to each different observer position. This figure shows that, for the majority of the galaxies, Eq.~(\ref{Rerhalf}) underestimates the stellar mass effective radius, particularly for the smaller galaxies. The individual measurements are colour-coded by the inclination of the galaxy for the given observer, defined as the angle between the direction of the stellar angular momentum vector and the direction towards the observer. For a given half-mass radius, galaxies observed nearly edge-on turn out to have, on average, smaller values for $R_{\text{e}}^\star$ than galaxies observed nearly face-on. The solid yellow line is a power-law fit to the individual data points and is given by
\begin{equation}
\frac{R_{\text{e}}^\star}{\text{kpc}} = 0.999 \left( \frac{r_{\text{half}}}{\text{kpc}}\right)^{0.922}.
\end{equation}
Most importantly, we have, for each galaxy and for each observer's position, a measurement of the half-mass radius determined in the exact same way as the half-light radius in each of the broadband filters. 


\section{Results}
\label{Results.sec}

\subsection{Wavelength dependence of the effective radius}
\label{wavdep.sec}

\begin{figure*}
\centering
\includegraphics[width=0.85\textwidth]{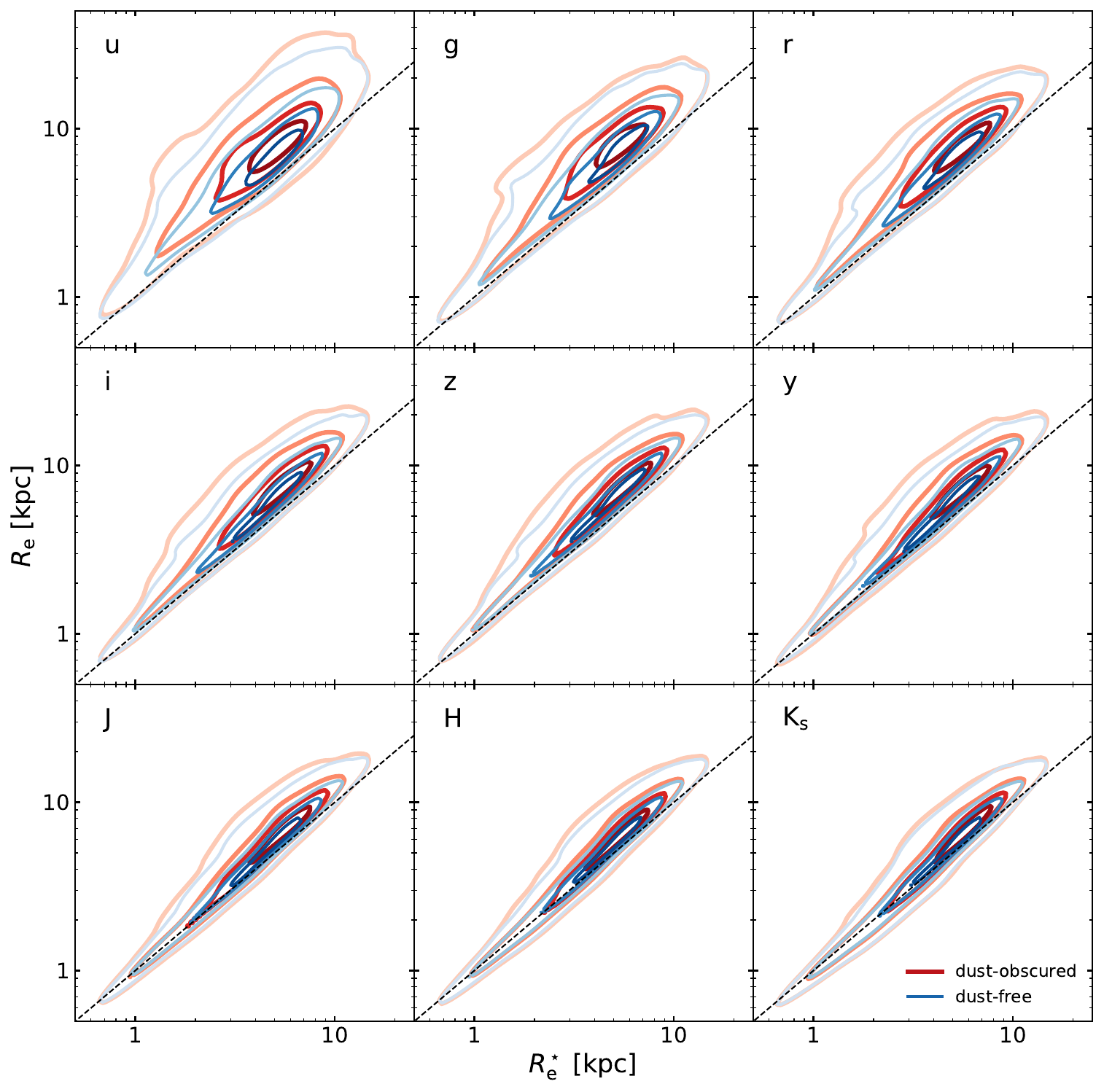}\hspace*{1em}
\caption{Comparison between the half-mass radius $R_{\text{e}}^\star$ and the half-light radius $R_{\text{e}}$ in nine different optical--NIR broadband filters. Thick red contours correspond to measurements of $R_{\text{e}}$ from the dust-obscured images, slightly thinner blue contours to measurements from dust-free images. The contour correspond to the 10, 40, 70 and 90\% percentile levels. The dashed lines represent a one-to-one relationship between $R_{\text{e}}^\star$ and $R_{\text{e}}$.}
\label{compare_sizes.fig}
\end{figure*}

In Figure~{\ref{compare_sizes.fig}} we compare the effective radius $R_{\text{e}}$ in each of the nine considered bands with respect to the half-mass radius $R_{\text{e}}^\star$. Focusing first on the blue contours, which correspond to the effective radii obtained from the dust-free images, we see that the dust-free effective radius is larger than the half-mass radius. This excess is largest in the {\it{u}}-band and systematically decreases towards the ${\textit{K}}_{\text{s}}$-band. Looking at the thicker red contours, which correspond to the images with dust attenuation properly accounted for, we note similar but even more outspoken trends. The effective radius as measured from the dust-obscured images systematically exceeds both the half-mass radius and the dust-free effective radius. The excess decreases with increasing wavelength.

\begin{figure}
\centering
\includegraphics[width=\columnwidth]{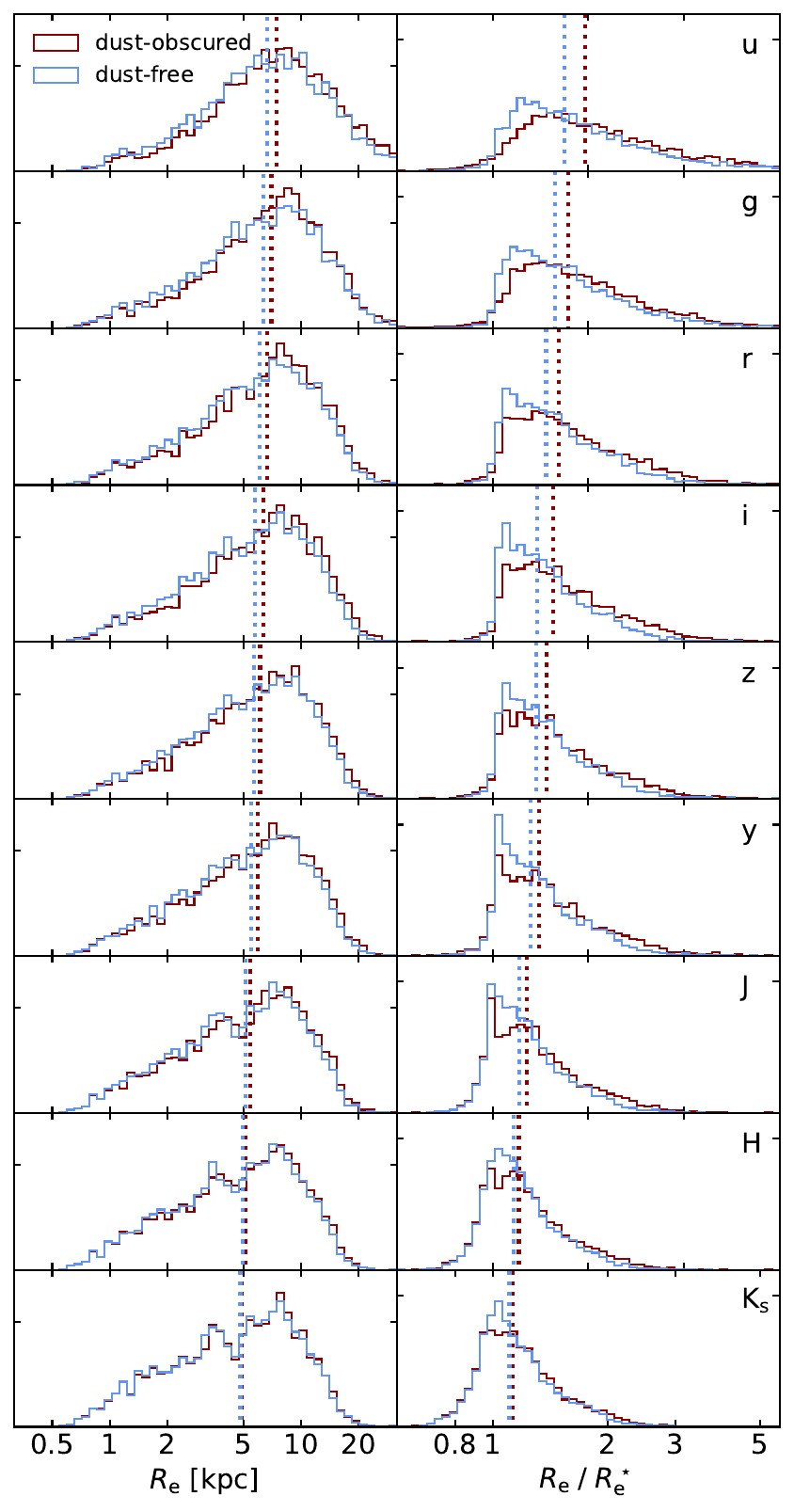}%
\caption{Histograms of $R_{\text{e}}$ and $R_{\text{e}}/R_{\text{e}}^\star$ for nine different optical--NIR broadband filters, for both dust-obscured (red) and dust-free (blue) images. The median values of the distributions are indicated as vertical dotted lines.}
\label{Re_histograms.fig}
\end{figure}

Figure~{\ref{Re_histograms.fig}} shows the same data in an alternative way. The histograms on the left column display a wide range of effective radii in any broadband filter, which just reflects the variety of sizes in the galaxy population. In each individual band, the distribution of effective radii approximates a lognormal distribution, similar to the effective radius distributions at low redshift shown by \citet{2014MNRAS.441.1340V} and at higher redshift by \citet{2015ApJS..219...15S}. Our distributions tend to be slightly skewed with a long tail towards smaller effective radii, whereas the distributions presented by \citet{2014MNRAS.441.1340V} are rather skewed towards larger effective radii.

More remarkable than the large spread in $R_{\text{e}}$ is that also the range in $R_{\text{e}}/R_{\text{e}}^\star$ is relatively broad in all the bands. At short wavelengths, this is not unexpected, as blue light is a poor estimator for the stellar mass. We find that the width of the distribution in $R_{\text{e}}/R_{\text{e}}^\star$ decreases when moving from shorter to longer wavelengths. Interestingly, this width is still non-negligible even in the ${\textit{K}}_{\text{s}}$-band that is generally considered as a good proxy for stellar mass. The ratio between the ${\textit{K}}_{\text{s}}$-band half-light ratio and half-mass ratio can be as small as 0.8 and as large as 2. 

\begin{figure}
\centering
\includegraphics[width=0.9\columnwidth]{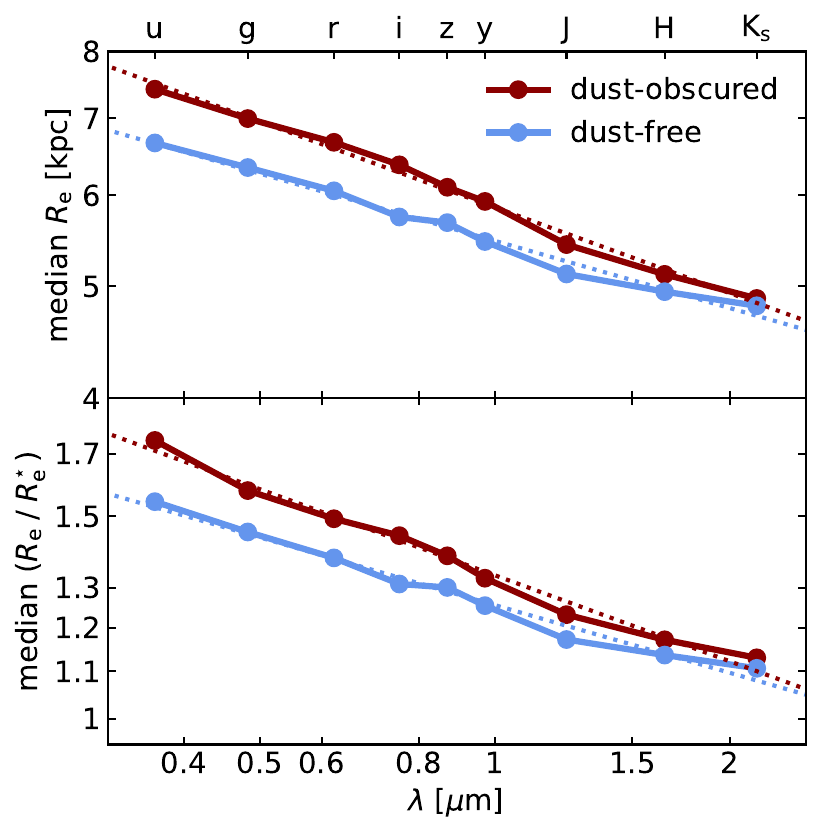}\hspace*{2em}
\caption{Wavelength dependence of the effective radius for our sample of TNG50 galaxies. Top panel: median value of the half-light radius. Middle panel: median ratio of the half-light radius to the {\it{g}}-band half-light radius. Bottom panel: median ratio of the half-light radius and half-mass radius. In all three panels, the black lines and markers correspond to the dust-affected images, the blue ones to the dust-free images. The dotted lines represent power-law fits to the data points.}
\label{sizeratio_wavelength.fig}
\end{figure}

Another interesting characteristic of these histograms is the systematic shift with increasing wavelength. Moving from top to bottom, one notes that the histograms of both $R_{\text{e}}$ and $R_{\text{e}}/R_{\text{e}}^\star$ slightly but systematically shift towards smaller values. This shift is present for both the dust-obscured and dust-free measurements of the effective radius. The systematic shift with increasing wavelength can be characterised by the median values of the histograms, indicated as dotted vertical lines in Figure~{\ref{Re_histograms.fig}}. One can immediately note that these median values systematically decrease as a function of increasing wavelength. 

In Figure~{\ref{sizeratio_wavelength.fig}} we show the median values of $R_{\text{e}}$ and $R_{\text{e}}/R_{\text{e}}^\star$ explicitly as a function of wavelength. In each panel, the red lines and markers correspond to the dust-obscured images, the blue ones to the dust-free images. These plots confirm the general trends noticeable in Figures~{\ref{compare_sizes.fig}} and {\ref{Re_histograms.fig}} and allow to make more quantitative observations. It shows that, across the optical--NIR range, the half-light radius tends to exceed the half-mass radius, and the median effective radius systematically decreases with increasing wavelength. In the optical {\textit{g}}-band the median ratio of $R_{\text{e}}/R_{\text{e}}^\star$ is 1.58, in the ${\textit{K}}_{\text{s}}$-band the median ratio is still 1.13. The two panels of Figure~{\ref{sizeratio_wavelength.fig}} also show that the effective radii measured from dust-obscured images are systematically larger than those measured from dust-free images. The {\textit{g}}-band half-light radii measured from dust-obscured images are, on average, 8.7\% larger than those measured from dust-free images. In the ${\textit{K}}_{\text{s}}$-band the median ratio of this excess reduces to 2.1\%.

\subsection{Origin of the wavelength dependence} 

These findings confirm the common understanding that the decrease in the effective radius of galaxies with increasing wavelength is the combination of two distinct effects: stellar population gradients and dust attenuation. We note that, in the present study, we do not make an explicit distinction between bulge and disc components. The half-light radius we measure is determined using a curve-of-growth analysis and is a nonparametric measure of the galaxy size. Any variation in the relative intrinsic luminosity of bulge and disc components as a function of wavelength is incorporated in our stellar population gradient.

The stellar population gradient contribution is evident from the fact that the effective radius measured from the dust-free images shows this systematic wavelength dependence. It can only be caused by differences in the intrinsic colours of the stellar populations. Differential dust attenuation and the corresponding flattening of the surface brightness profiles is responsible for an additional contribution to the wavelength dependence of the half-light radius.

\begin{table}
\caption{Parameters of power-law fits to the median wavelength dependence of the effective radius, $\log X = \alpha \log(\lambda/\mu{\text{m}}) + \beta$. The second and third columns correspond to dust-obscured data, the fourth and fifth column to dust-free data. }
\label{fits.tab}
\centering
\begin{tabular}{crrrr}
$X$ & $\alpha$\hspace*{1em} & $\beta$\hspace*{1em}  & $\alpha_{\text{df}}$\hspace*{1em}  & $\beta_{\text{df}}$\hspace*{1em}  \\[0.5em] \hline \\[-0.2em]
$R_{\text{e}}/{\text{kpc}}$ & $-0.248$ & $0.767$ & $-0.194$ & $0.739$ \\
$R_{\text{e}}/R_{\text{e}}^{\star}$ & $-0.249$ & $0.125$ & $-0.196$ & $0.099$ \\[0.5em] \hline
\end{tabular}
\end{table}

Contrary to purely observational imaging studies \citep[e.g.][]{2005JKAS...38..149K, 2010MNRAS.408.1313L, 2012MNRAS.421.1007K, 2014MNRAS.441.1340V}, we can try to disentangle the importance of the two effects to the observed wavelength dependence of the half-light radius because we can artificially turn dust attenuation on or off. We fitted power-law fits to the median trends in Figure~{\ref{sizeratio_wavelength.fig}}, shown as the dotted lines in each panel. The parameters of the fits are listed in Table~{\ref{fits.tab}}. While the two contributions (stellar population gradients and differential dust attenuation) do not linearly add up, we can take the slope of the power-law as a qualitative measure of the importance of each contributor. The joint contribution of stellar population gradients and differential dust attenuation results in a slope of almost $-0.25$ for the two power-law fits. When we consider only stellar population gradients, that is, when we suppress dust attenuation, we find a slope of nearly $-0.20$ for the two power-law fits. These numbers suggest that the stellar population gradients are the dominant factor in driving the wavelength dependence of the effective radius, and that differential dust attenuation is a minor, but non-negligible, secondary factor. In the concise style of \citet{2004ApJ...600L..39T} we can summarise the wavelength dependence of galaxy sizes: 80\% stellar population, 20\% dust attenuation.


\section{Discussion}
\label{Discussion.sec}

\subsection{Comparison to Vulcani et al.\ (2014)}
\label{comparisonlit.sec}

\begin{figure}
\centering
\includegraphics[width=0.9\columnwidth]{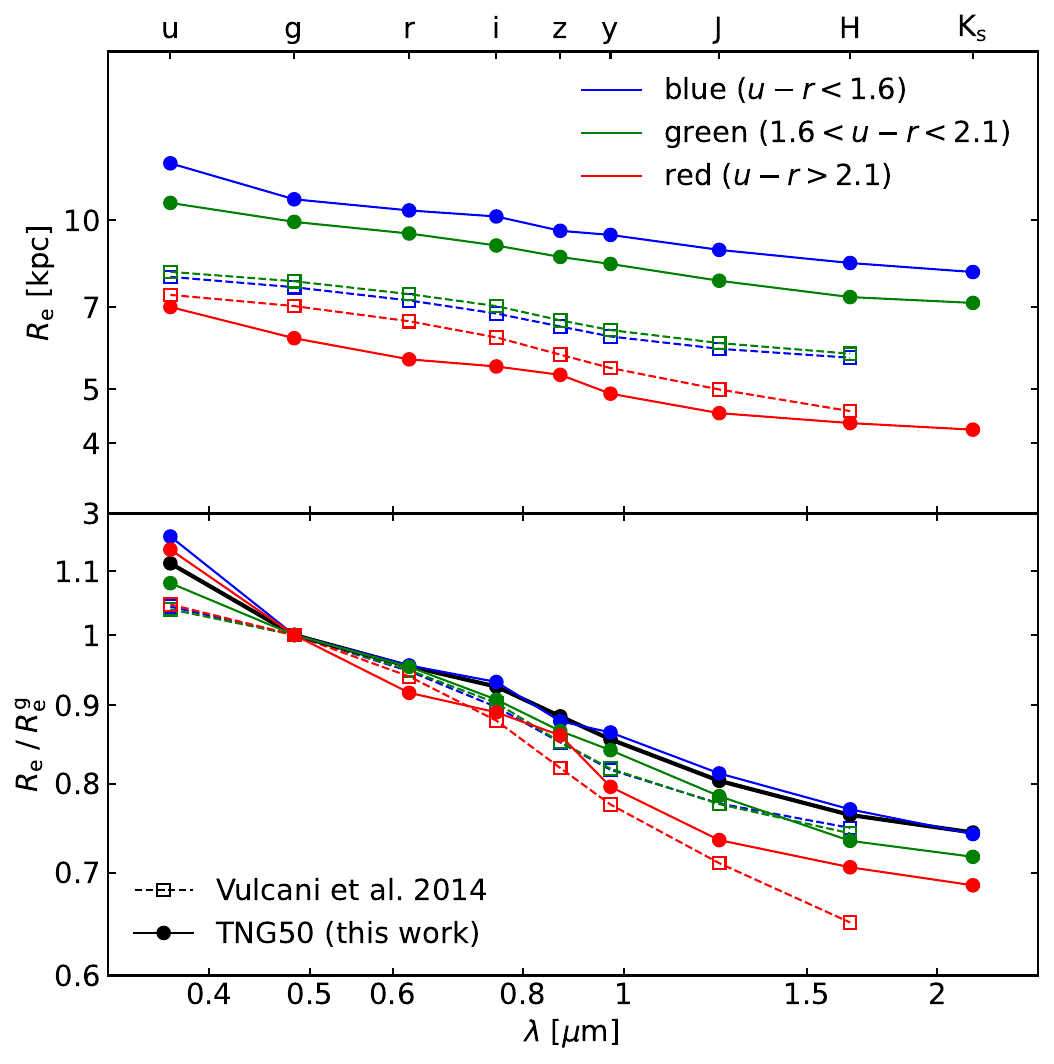}\hspace*{1em}
\caption{Comparison of the wavelength dependence of the effective radius for our sample of TNG50 galaxies to literature values from \citet{2014MNRAS.441.1340V}. Top panel: median value of the half-light radius. Bottom panel: median ratio of the half-light radius to the {\it{g}}-band half-light radius. For the meaning of the different colours, we refer to Sect.~{\ref{comparisonlit.sec}}. }
\label{vulcani.fig}
\end{figure}

The results of the previous section are only valid if the TNG50 galaxies in our sample are representative for the galaxies in the local Universe. In this subsection we compare the wavelength dependence of the effective radius of our TNG50 galaxies with the work by \citet{2014MNRAS.441.1340V}. These authors presented the wavelength dependence of the effective radius based on a large sample of galaxies with available optical (SDSS {\it{ugriz}}) and NIR (UKIDSS {\it{YJHK}}) imaging from the Galaxy And Mass Assembly \citep[GAMA,][]{2011MNRAS.413..971D} survey.\footnote{The SDSS {\textit{ugriz}} + UKIDDS {\textit{YJHK}} filter set is slightly different from the LSST {\textit{ugrizy}} + 2MASS {\textit{JHK}}$_{\text{s}}$ filter set that we used for our SKIRT post-processing, but this does not fundamentally affect the conclusions of the comparison.} The authors considered a complete, volume-limited sample with an absolute {\textit{r}}-band magnitude limit of $M_r = -21.2$. They split their sample of galaxies into three different populations based on ${\textit{u}}-{\textit{r}}$ colour: blue (${\textit{u}}-{\textit{r}} < 1.6$), green ($1.6 < {\textit{u}}-{\textit{r}} < 2.1 $), and red (${\textit{u}}-{\textit{r}} > 2.1$). They simultaneously fitted one wavelength-dependent S\'ersic model to the optical and NIR images with the GALFITM code \citep{2013MNRAS.430..330H}. The dashed lines and open squares in Figure~{\ref{vulcani.fig}} represent the median values for $R_{\text{e}}$ and $R_{\text{e}}/R_{\text{e}}^{\text{g}}$ as a function of wavelength obtained by \citet{2014MNRAS.441.1340V} for their three subpopulations.

We attempted to mimic the sample selection criteria of \citet{2014MNRAS.441.1340V} by excluding from our sample all the galaxies with $M_r>-21.2$ and subsequently subdividing the resulting sample into three subpopulations using the same ${\textit{u}}-{\textit{r}}$ colour criteria. The resulting median values for $R_{\text{e}}$ and $R_{\text{e}}/R_{\text{e}}^{\text{g}}$ are shown as the solid lines and filled circles in Figure~{\ref{vulcani.fig}}.

The top panel of Figure~{\ref{vulcani.fig}} shows that there is a significant difference between the absolute values of the effective radii for the different populations. For the red population, there is a reasonable agreement, with our values, averaged over the entire wavelength range, some 9\% smaller than the values by \citet{2014MNRAS.441.1340V}. For the green and blue populations, however, there is a significant difference: the median effective radii for our green population are about 29\% larger than theirs, and for the blue population this increases to almost 50\%.

We see three potential reasons that can contribute to these significant differences. The first reason is we used different techniques to determine the effective radii. \citet{2014MNRAS.441.1340V} use single-component S\'ersic fits to the images. Such fitting simultaneously determines the total magnitude, effective radius, and S\'ersic index as parameters of the parametric model that provides the best description of the surface brightness distribution in a $\chi^2$ fitting. Our half-light radii, on the other hand, are determined using a curve-of-growth analysis (see Sect.~{\ref{measurement.sec}}). It is well-known that different measures for the effective radius can differ systematically, as also demonstrated by \citet{2014MNRAS.441.1340V} in their Figure~1. Moreover, the spatial resolution (i.e.\ PSF size and pixel size) of the observed imaging data may also have a significant effect on the observed trends. 

A second aspect that can at least be partly explain the differences are sample selection affects. While we adopted the same absolute magnitude and colour criteria, the sample considered by \citet{2014MNRAS.441.1340V} contains galaxies with redshifts up to $z=0.3$, with a peak in the redshift distribution around 0.2. We only have access to simulated galaxies at $z=0$ as the TSA currently only considers this single TNG50 snapshot. Since galaxy properties, including galaxy sizes, do change as a function of cosmic time \citep[e.g.][]{2014ARA&A..52..291C}, we cannot expect to have a perfect match at this stage. 

Finally, we cannot exclude that the TNG50 galaxies have systematically larger sizes than the galaxies considered by \citet{2014MNRAS.441.1340V}. Checking this is not straightforward without a detailed matched sample. We come back to this issue in Sect.~{\ref{SGA-2020.sec}}.

One would hope that sample selection effects and the differences between the methods used to measure $R_{\text{e}}$ would be less important when considering the systematic wavelength dependence of the effective radius rather than its absolute value. This seems indeed the case, as shown in the bottom panel of Figure~{\ref{vulcani.fig}}. An interesting feature of this plot is the difference between the red population on the one hand, and the green--blue population on the other hand. The wavelength dependence of the former population is steeper than the one of the latter, both for the observed galaxies and for the simulated TNG50 galaxies. The wavelength dependence of $R_{\text{e}}$ is due to a combination of stellar population gradients and differential dust attenuation, but as red, bulge-dominated galaxies are generally less dusty than blue, disc-dominated galaxies \citep[e.g.][]{2010MNRAS.403.1894D, 2012A&A...540A..52C, 2012ApJ...748..123S, 2019A&A...624A..80N}, this difference in slope must be caused by stronger stellar population gradients. This result is in agreement with our conclusion from Sect.~{\ref{wavdep.sec}} that stellar population gradients are the dominant factor in driving the wavelength dependence of the effective radius.

\subsection{Comparison to the Siena Galaxy Atlas}
\label{SGA-2020.sec}

\begin{figure}
\centering
\includegraphics[width=0.9\columnwidth]{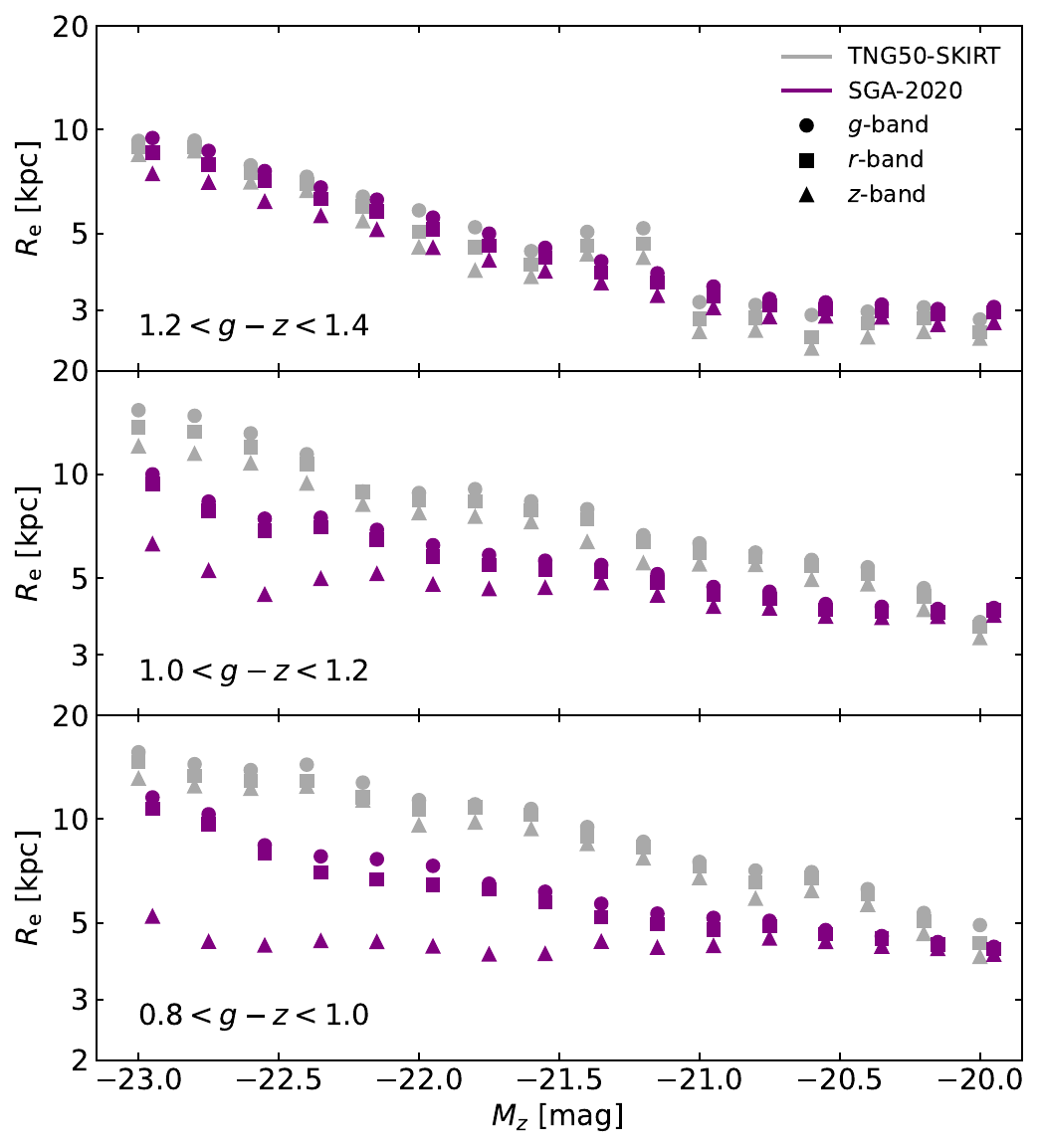}\hspace*{1em}
\caption{Comparison of the effective radii for our sample of simulated TSA galaxies to effective radii of galaxies from the Siena Galaxy Atlas 2020 \citep{2023ApJS..269....3M}. The three different panels correspond to different bins in ${\textit{g}}-{\textit{z}}$ colour, and each symbol represents the median effective radius in a bin in {\textit{z}}-band absolute magnitude of width $\Delta M_z = 0.2$. Measurements in the {\textit{g}}-, {\textit{r}}-, and {\textit{z}}-bands are indicated by different symbols, as indicated in the top panel. The symbols for the SGA-2020 are plotted slightly offset for the sake of clarity.}
\label{SGA-2020.fig}
\end{figure}

\begin{figure*}
\centering
\includegraphics[width=\textwidth]{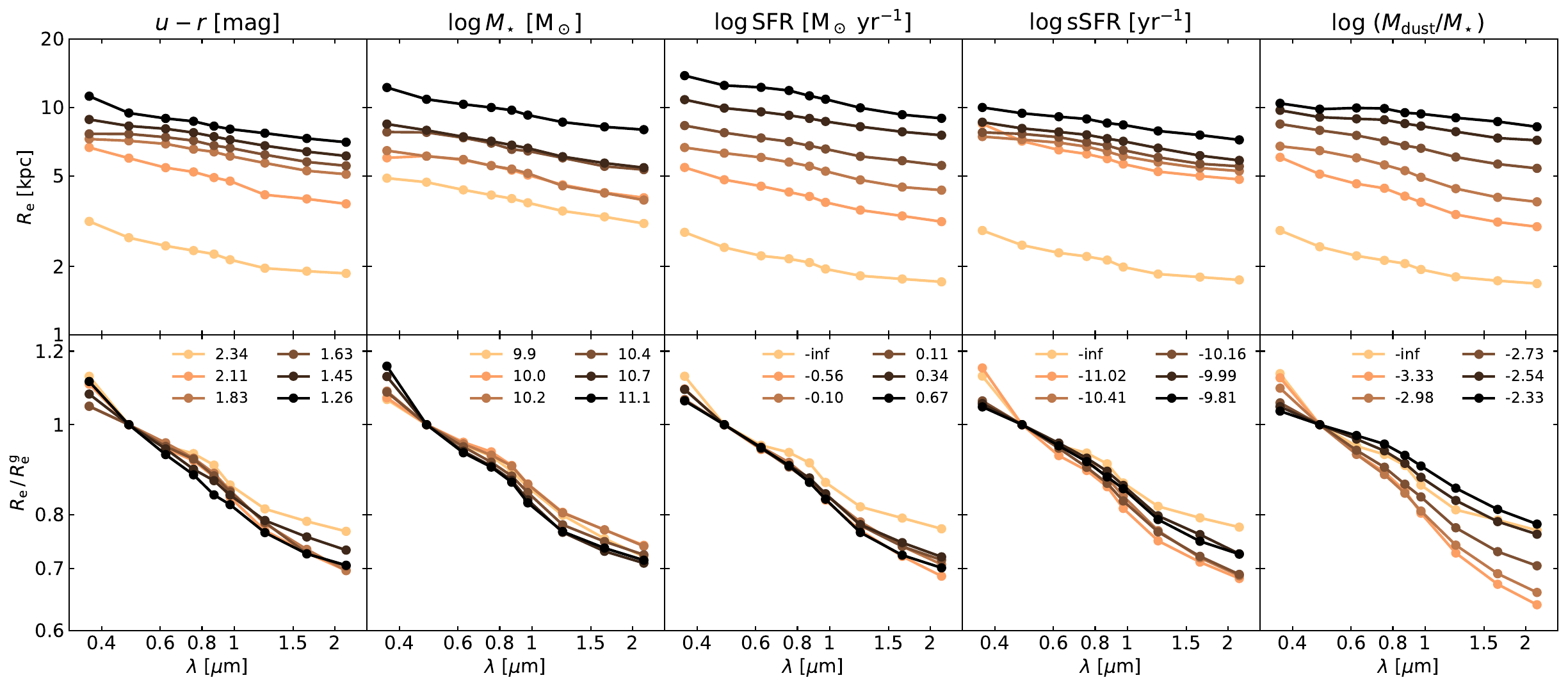} \\[0.5em]
\hspace*{0.2em}\includegraphics[width=0.996\textwidth]{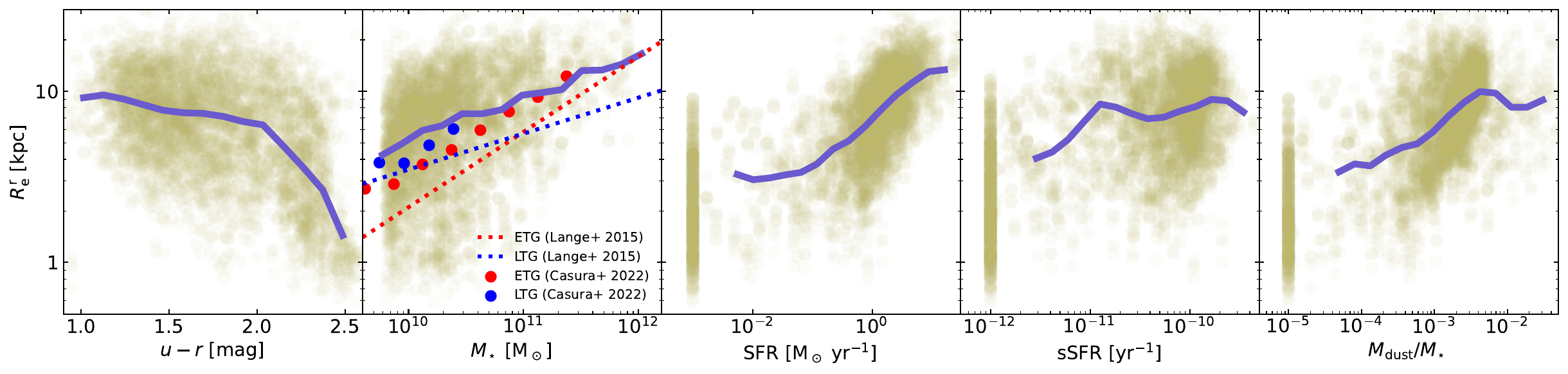}
\caption{Dependence of the effective radius on global physical properties. Each column corresponds to a different global physical property: ${\textit{u}}-{\textit{r}}$ colour, stellar mass, star formation rate, specific star formation rate, and specific dust mass. The entire sample is split into six subpopulations based on each of these properties. Top row: median effective radius for each subpopulation. Middle row: median ratio of the effective radius and the {\textit{g}}-band effective radius. Bottom row: scatter plots relating the physical property to the {\textit{g}}-band effective radius. All TNG50 galaxies without ongoing star-formation are plotted at ${\text{SFR}} = 10^{-3}~{\text{M}}_\odot~{\text{yr}}^{-1}$ and ${\text{sSFR}} = 10^{-12}~{\text{yr}}^{-1}$, and all dust-free galaxies at $M_{\text{dust}}/M_\star = 10^{-5}$. The blue curves represent running medians.}
\label{sizeratio_wavelength_split.fig}
\end{figure*}

As indicated in the previous section, we found a significant difference in the absolute values of the effective radii of our TSA galaxies and the observed galaxies considered by \citet{2014MNRAS.441.1340V}. To investigate to which degree the sample selection is responsible for these differences, we compare the effective radii of our TSA galaxies to effective radii for galaxies from the 2020 version of the Siena Galaxy Atlas (SGA-2020 or simply SGA) presented by \citet{2023ApJS..269....3M}. The SGA is an angular-size-selected multi-wavelength atlas of almost 400,000 nearby galaxies (out to $z<0.5$). The catalogue is based on deep, wide-field {\textit{grz}} imaging from the DESI Legacy Imaging Surveys DR9 \citep{2019AJ....157..168D}. 

For the vast majority of the galaxies in the SGA, the effective radius in each of the {\textit{grz}}-bands is determined using a method that is very similar to the method we applied. Elliptical isophotes are fitted to the images and the curve-of-growth is determined by integrating the flux within the subsequent elliptical apertures. The main difference to our approach is that the curve-of-growth is fitted by an empirical model from which the effective radius is determined, whereas we use a spline interpolation.

Compared to the multi-wavelength study by \citet{2014MNRAS.441.1340V}, which showed the wavelength dependence of the effective radius in 8 bands from {\textit{u}} to {\textit{H}}, the SGA only lists effective radii in three optical bands. On the other hand, this sample has the advantages that we can construct galaxy samples that match more closely, and that the methods used to measure the half-light radius are very similar.

We started our selection by picking from the SGA sample all galaxies with distances between 50 and 150~Mpc. By eliminating galaxies with very low redshifts we ensure that the redshifts are dominated by the Hubble expansion and not by peculiar velocities; by eliminating galaxies with high redshifts we avoid a non-negligible effect of K-corrections and small angular sizes that might make the determination of the effective radius from the DESI imaging data more difficult. We further eliminate all galaxies without measurements of the effective radius in all three bands, we converted the model magnitudes to absolute magnitudes, and we calculated the effective radii in physical units. 

To generate a matched sample, we subdivided the observed SGA and simulated TSA samples into two-dimensional bins in {\textit{z}}-band absolute magnitude and ${\textit{g}}-{\textit{z}}$ colour. We limited the sample to $ -23.1 < M_z < -19.9$ and $0.8 < {\textit{g}}-{\textit{z}} < 1.4$ to ensure sufficient galaxies in each bin. We considered 16 bins in absolute magnitude with width $\Delta M_z = 0.2$ and three bins in ${\textit{g}}-{\textit{z}}$ colour with $\Delta({\textit{g}}-{\textit{z}}) = 0.2$. In each bin we calculated the median effective radii of the two galaxy samples in the {\textit{g}}-, {\textit{r}}-, and {\textit{z}}-bands. The comparison is shown in Figure~{\ref{SGA-2020.fig}}.

In all 48 bins, we find that the effective radius decreases from the ${\textit{g}}$- to the ${\textit{r}}$- to the ${\textit{z}}$-band. This is in line with the results from our analysis, namely that the effective radius systematically decreases with increasing wavelength. For the reddest galaxies, i.e. with $1.2 < {\textit{g}}-{\textit{z}} < 1.4$, we find excellent agreement between the effective radii of the SGA-2020 and the TSA galaxies. The median {\textit{r}}-band effective radius systematically decreases from almost 9~kpc for $M_z=-22$ to 3~kpc for $M_z=-20$. For bluer galaxies with ${\textit{g}}-{\textit{z}} < 1.2$, we find significant differences (the strong decrease in the effective radius between {\textit{r}}-band and {\textit{z}}-band for the most luminous blue galaxies in the SGA seems spurious). For both the SGA and the TSA samples, the median effective radius systematically decreases for decreasing luminosity, but the absolute values differ. For the fainter galaxies with $M_z\approx-20$ the median effective radii agree fairly well, but for more luminous galaxies the TSA galaxies are systematically larger than the observed SGA galaxies. The largest differences are found for intermediate luminosities ($M_z \approx -21.5$) and for the bluest colours $(0.8 < {\textit{g}}-{\textit{z}} < 1.0$): in this case the median {\textit{r}}-band effective radius of simulated TSA galaxies exceeds the corresponding value of the observed SGA galaxies by about 70\%.

Having constructed closely matched samples in which the effective radii are measured in very similar ways, we conclude that difference in median effective radius for the bluer galaxies is most probably due to intrinsic differences in the morphological structure of observed and TNG50 simulated galaxies. On the one hand, this seems surprising, as the IllustrisTNG model is by construction designed to match galaxy sizes at low redshift \citep[][Figs.~4 and 8]{2018MNRAS.473.4077P}. In their morphological comparison of TNG100 simulated galaxies and Pan-STARRS observed galaxies, \citet{2019MNRAS.483.4140R} measured systematically larger half-light radii at the massive end ($M_\star > 10^{11}~{\text{M}}_\odot$), while the apparent sizes of less massive galaxies agree very well with the observations. On the other hand, the IllustrisTNG model is calibrated at the resolution of the TNG100 simulation, and is known to generate slightly increased stellar masses and SFRs at the increased resolution of the TNG50 simulation \citep{2018MNRAS.473.4077P, 2018MNRAS.475..648P, 2019MNRAS.490.3196P, 2019MNRAS.485.4817D, 2022MNRAS.516.3728T}. Using intrinsic half-mass radii extracted from the TNG public database, we find that the median half-mass radius of TNG50 galaxies in the stellar mass range we consider ($10^{9.8}~{\text{M}}_\odot < M_\star < 10^{12}~{\text{M}}_\odot$) is 20\% larger than that of TNG100 galaxies in the stellar same mass range (4.55~kpc versus 3.73~kpc).

In future work we will apply different techniques to investigate the morphological structure of the TSA galaxies, including both parametric modelling (one-component S\'ersic fitting and bulge-disc decomposition) and nonparametric methods. 

\subsection{Dependence of the effective radius on global physical properties}

In the previous two subsections, we separated our galaxy sample in different subpopulations according to ${\textit{u}}-{\textit{r}}$ colour (Sec.~{\ref{comparisonlit.sec}}) and {\textit{z}}-band absolute magnitude and ${\textit{g}}-{\textit{z}}$ colour (Sec.~{\ref{SGA-2020.sec}}). In this subsection we want to generalise these results: we investigate how the absolute value of $R_{\text{e}}$ and the wavelength dependence of $R_{\text{e}}$ vary when we split our sample into different subpopulations by means of different global physical properties. We only use the half-light radii derived from the dust-obscured images in this subsection.

The result of this analysis is shown in Figure~{\ref{sizeratio_wavelength_split.fig}}. Each column in this figure corresponds to a different global physical property: ${\textit{u}}-{\textit{r}}$ colour, stellar mass, star formation rate, specific star formation rate, and dust-to-stellar-mass ratio (specific dust mass). The ${\textit{u}}-{\textit{r}}$ colour was calculated from the SKIRT-generated dust-obscured images, while the other properties were calculated directly from the TNG50 simulation particle data. For each of these five properties, we split the galaxy sample into six bins with the same number of galaxies in each bin. For the median value of the physical property per bin, see the legend in Figure~{\ref{sizeratio_wavelength_split.fig}}, middle row. For each bin, we calculate the median value of $R_{\text{e}}$ (top row) and $R_{\text{e}}/R_{\text{e}}^{\text{g}}$ (middle row) in every band. 

The bottom row shows a scatter plot of the {\textit{r}}-band half-light radius against each of the global physical properties. The solid blue curves represent running medians for our sample. For the scatter plot that shows the effective radius versus the stellar mass, i.e.\ the size--mass relation, we also show results from two observational studies. The dotted lines show the linear fits to the size--mass relation of early- and late-type galaxies from \citet{2015MNRAS.447.2603L}, where the S\'ersic index is used to discriminate between the two classes. The filled circles represent the mean values for the size-mass relation from \citet{2022MNRAS.516..942C} for the same two populations. The results are largely compatible with our trends, although at fixed stellar mass, the median effective radii from our TSA galaxies tend to exceed the observational relations, in agreement with the findings from Sect.~{\ref{SGA-2020.sec}}. A full analysis of the size--mass relation of the galaxies in the TSA, including the dependence on secondary galaxy parameters and the evolution with redshift, is beyond the scope of this paper and will be the topic of future work. For the present paper, the scatter plots at the bottom row of Figure~{\ref{sizeratio_wavelength_split.fig}} mainly serve to interpret the results in the top two rows.

The panels in the first column expand upon the results from Figure~{\ref{vulcani.fig}} by now splitting the sample into six different bins in {\textit{u}}--{\textit{r}} colour. At every wavelength, the median effective radius decreases with increasing ${\textit{u}}-{\textit{r}}$ colour. The reddest galaxies in our sample with ${\textit{u}}-{\textit{r}}>2.1$ are more compact than bluer galaxies with ${\textit{u}}-{\textit{r}}<2.1$. This behaviour can be understood by looking at the bottom left panel. The reddest galaxies are very compact: there are virtually no large galaxies with  ${\textit{u}}-{\textit{r}} > 2.4$. Concerning the slope of the wavelength dependence of $R_{\text{e}}$ our results are more mixed: there does not seem to be a systematic and consistent change with increasing colour.

In general, we find a similar behaviour with other global physical parameters. Splitting the galaxies in different subpopulations, we find a systematic change of the median half-light radius at every wavelength. More specifically, the median effective radius in every single band increases with galaxy stellar mass, total SFR, sSFR, and dust-to-stellar-mass ratio. In particular for SFR and specific dust mass, the different subpopulations are clearly separated in the panels on the top row, with positive correlations also clearly visible in the panels on the bottom row. 

There does not seem to be a systematic correlation with any global property for the slope of the wavelength dependence of $R_{\text{e}}$. There is hardly any difference in the median slope for the different subpopulations in stellar mass, SFR, or sSFR. When we subdivide the sample according to specific dust mass, we do find different median slopes, but they do not increase or decrease systematically or monotonically over the entire parameter range. 

We interpret this lack of a monotonic correlation between the slope of the wavelength dependence of galaxy sizes and any physical property as the result of the two different effects that contribute to this wavelength dependence: stellar population gradients and dust attenuation (Sect.~{\ref{wavdep.sec}}). The mean metallicity and age gradients in both late-type and early-type galaxies vary strongly with galaxy morphology, galaxy stellar mass and environment \citep[e.g.][]{2010MNRAS.407..144T, 2017MNRAS.465.4572Z, 2019MNRAS.483.1862Z}. On the other hand, also specific dust masses and the level of attenuation depends strongly on galaxy morphology and stellar mass \citep{2012A&A...540A..52C, 2012ApJ...748..123S, 2013MNRAS.428.1880A, 2016A&A...586A..13V, 2018A&A...620A.112B, 2019A&A...624A..80N}. It is hence not surprising that we find no systematic correlation between the slope of the $R_{\text{e}}(\lambda)$ relation and a single global galaxy parameter.


\section{Summary and conclusion}
\label{Conclusions.sec}

Based on the TSA, a multi-wavelength atlas of high-resolution synthetic images of simulated galaxies extracted from the TNG50 cosmological simulation, we investigated the wavelength dependence of the effective radius of galaxies at optical and NIR wavelengths. We determined the half-light radius in each individual image (and each stellar mass surface density map) using a curve-of-growth analysis. The main results we obtained are the following:
\begin{itemize}
\item The effective radius is, in every band, on average larger than the half-mass radius. This excess is largest in the {\textit{u}}-band and systematically decreases with increasing wavelength. In the optical ${\textit{g}}$-band the median ratio of $R_{\text{e}}/R_{\text{e}}^\star$ is 1.58, in the ${\textit{K}}_{\text{s}}$-band the median ratio is still 1.13. 
\item Effective radii measured from dust-obscured images are systematically larger than those measured from dust-free images. The {\textit{g}}-band half-light radii measured from dust-obscured images are, on average, 8.7\% larger than those measured from dust-free images. In the ${\textit{K}}_{\text{s}}$-band the median ratio is almost negligible at 2.1\%.
\item The decrease in the effective radius of galaxies with increasing wavelength is due to the combination of two distinct effects: stellar population gradients and dust attenuation. Since we have both dust-obscured and dust-free images, we can disentangle the importance of the two effects. Assuming that our recipe to populate the galaxies with dust is appropriate, we find that stellar population gradients are the dominant factor (about 80\%) in driving the wavelength dependence of the effective radius, followed by differential dust attenuation (20\%).
\item Comparing our results to the observational study by \citet{2014MNRAS.441.1340V}, we find significant offsets in the absolute values of the median effective radii. In particular, the median effective radii for our green and blue populations are about 30 to 50\% larger than the median values obtained by \citet{2014MNRAS.441.1340V}. We also compare the effective radii of our galaxies to those of a carefully matched sample of observed galaxies from the Siena Galaxy Atlas in different bins in {\textit{z}}-band absolute magnitude and ${\textit{g}}-{\textit{z}}$ colour. We find excellent agreement for the reddest galaxies, but significant differences for the bluer galaxies. The difference in median ${\textit{r}}$-band effective radius peaks at about 70\%. Since the samples match closely and the approaches used to determine the effective radius are very similar, we interpret these differences as due to intrinsic differences in the morphological structure of observed and TNG50 simulated galaxies.
\item Contrary to the absolute values of the effective radii, we find that the slope of the wavelength dependence of the half-light radius of our TSA galaxies agrees more closely with the observational data from \citet{2014MNRAS.441.1340V}. We recover the observational result that the mean wavelength dependence for the red galaxy population is steeper than that for the green--blue population. Stronger stellar population gradients are probably the cause of this difference.
\item We investigated how the absolute value of $R_{\text{e}}$ and the wavelength dependence of $R_{\text{e}}$ vary when we split our sample into different subpopulations by means of different global physical properties. We find that the median effective radius in every single band increases systematically with decreasing ${\textit{u}}-{\textit{r}}$ colour and with increasing galaxy stellar mass, total SFR, sSFR, and dust-to-stellar-mass ratio. For the slope of the wavelength dependence of $R_{\text{e}}$ there does not seem to be a systematic, monotonic correlation with any global property. We interpret this as the result of the complex dependence of both population gradients and attenuation on galaxy morphology, stellar mass and environment. 
\end{itemize}
As discussed in \citet{Baes2024a}, the TSA allows for a suite of applications, in particular concerning the link between intrinsic galaxy properties and galaxy morphology. In future papers we plan to use this image atlas to systematically investigate the wavelength dependence and the effects of dust attenuation on other morphological characteristics, derived from single or multiple-component S\'ersic fitting or from a non-parametric morphological analysis. We also envision to extend the TSA by adding similar images at different redshifts, which would allow a direct investigation of the cosmic evolution of galaxy morphology in the TNG50 simulation and direct apples-to-apples comparison to observational data. Specifically focusing on the effective radius discussed in this paper, it would shed light on the nature of the apparent size evolution of galaxies, the origin of which is still under discussion \citep[e.g.][]{2019ApJ...877..103S, 2023arXiv230703264V}.


\begin{acknowledgements}
The authors thank the anonymous referee for comments and suggestions that significantly improved the content and presentation of this work. MB, NA, IK, and MM acknowledge financial support by the Flemish Fund for Scientific Research (FWO-Vlaanderen) through the research projects G030319N, G0G0420N, and G037822N.  AG and BVM acknowledge the support by FWO-Vlaanderen through the PhD Fellowship Grants 11H2121N and 1193222N, respectively. CT acknowledges the INAF grant 2022 LEMON. This project has received funding from the European Research Council (ERC) under the European Union's Horizon 2020 research and innovation program (grant agreement No.\ 683184).
\end{acknowledgements}


\bibliography{mybib}

\end{document}